# SPONTANEOUS MUON EMISSION DURING FISSION A NEW NUCLEAR RADIOACTIVITY

**D. B. Ion[1,2], M. L. D. Ion[3] , Reveica Ion-Mihai[3]**

[1] *Institute for Physics and Nuclear Engineering, IFIN-HH, Bucharest Romania*

[1] *Academy of Romanian Scientist*

[2] *Bucharest University, Faculty of Physics, Bucharest, Romania*

**Abstract**

In this paper the essential theoretical predictions for the nuclear muonic radioactivity are presented by using a special *fission-like model* similar with that used in description of the pionic emission during fission. Hence, a fission-like model for the muonic radioactivity takes into account the essential degree of freedom of the system: $\mu-$ fissility, $\mu-$ fission barrier height, etc. Using this model it was shown that most of the SHE-nuclei lie in the region where the muonic fissility parameters attain their limiting value $X_{\mu F}=1$ . Hence, the SHE-region is characterized by the absence of a classical barrier toward spontaneous muon and pion emissions. Numerical estimations on the $\Gamma_{\mu}/\Gamma_{SF}$ for the natural muonic radioactivities of the transuranium elements as well numerical values for barrier heights are given only for even-even parent nuclei. Some experimental results from LCP-identification emission spectrum are reviewed. Also, the experimental results obtained by Khryachkov et al, using new spectrometer for investigation of ternary nuclear fission, are presented. The OPERA-experiment proposed to perform search for muonic radioactivity from lead nuclei, in the low background conditions offered by the Gran Sasso underground Laboratory (LNGS), is discussed.

## 1. Introduction

The traditional picture of the nucleus as a collection of neutrons and protons bound together via the strong force has proven remarkable successful in understanding a rich variety of nuclear properties. However, the achievement of modern nuclear physics that not only nucleons are relevant in the study of nuclear dynamics but that pions and the baryonic resonances like $\Delta$ 's and N* play an important role too. So, when the nucleus is explored at short distance scales the presence of short lived subatomic particles, such as the pion and delta, are revealed as nuclear constituents. The role of pions, deltas, quarks and gluons in the structure of nuclei is one of challenging frontier of modern nuclear physics. This modern picture of the nucleus brings us to the idea [1] to search for new exotic natural radioactivities such as: $(\pi, \mu, \Delta, N^*)$ -emission during the nuclear fission in the region of heavy and superheavy nuclei. So, in 1985, we initiated the investigation of the nuclear spontaneous pion emission as a new possible nuclear radioactivities called *nuclear pionic radioactivity* (NPIR) (see Refs [1-36] in our review paper [2]) with possible essential contributions to the instability of heavy and superheavy nuclei (see [16-19, 29] in Ref. [2]). Moreover, new exotic radioactivities such as new mode of nuclear fission (*muonic radioactivity* [3], $\Lambda^0-radioactivity$ [4], *hyperfision* [5], *deltonic fission*[6], etc), were also suggested and investigated.

In this paper the essential theoretical predictions for the nuclear muonic radioactivity are presented by using a special *fission-like model* similar with that used in description [2] of the pionic emission during fission. Hence, a fission-like model for the muonic radioactivity takes into account the essential degree of freedom of the system: $Q_{\mu^\pm}$ -liberated energy (see Fig.1), $\mu-$ fissility (see Fig.2), $\mu-$ fission barrier height (see Fig.3), numerical predictions for relative yields $\Gamma_{\mu}/\Gamma_{SF}$ (Fig.4), are presented in Sect.2. Some examples of the experimental LCP-identification spectra, characteristic of light charged particle emission (LCP) during fission, are presented in Figs. 5-6 from Sect. 3, The supergiant halos produced by charged $(\mu^{\pm}+\pi^{\pm})-$ mesons (see fig.7), as well as, the proposed OPERA-experiment for the muonic radioactivity of Pb, are also discussed in Sect. 3. The Sect.4 is reserved for summary and conclusions.

## 2. Investigation of the pair (muon, neutrino)-emission during spontaneous fission.

### 2.1 Energy Liberated in Muonic Radioactivity

We started our investigations with the values of energies liberated in n-body fission of heavy nuclei.

$$(A, Z) \rightarrow \mu^{\pm} + \bar{\nu}_{\mu}(\nu_{\mu}) + (Z_1, A_1) + (Z_2, A_2)$$

$$(A, Z) \rightarrow \mu^{\pm} + \bar{\nu}_{\mu}(\nu_{\mu}) + (A_1, Z_1) + (A_2, Z_2) + \ldots .. + (A_n, Z_n) \quad \text{(prompt muons)}$$

Therefore, the energy liberated in spontaneous muon-neutrino pair accompanied by two-body fission is defined as in usual case of a nuclear reaction, by the relation:

$$Q_{\mu^{\pm}} \equiv Q_{\mu F} = M_A - M_{A_1} - M_{A_2} - m_{\mu} - m_{\nu}$$

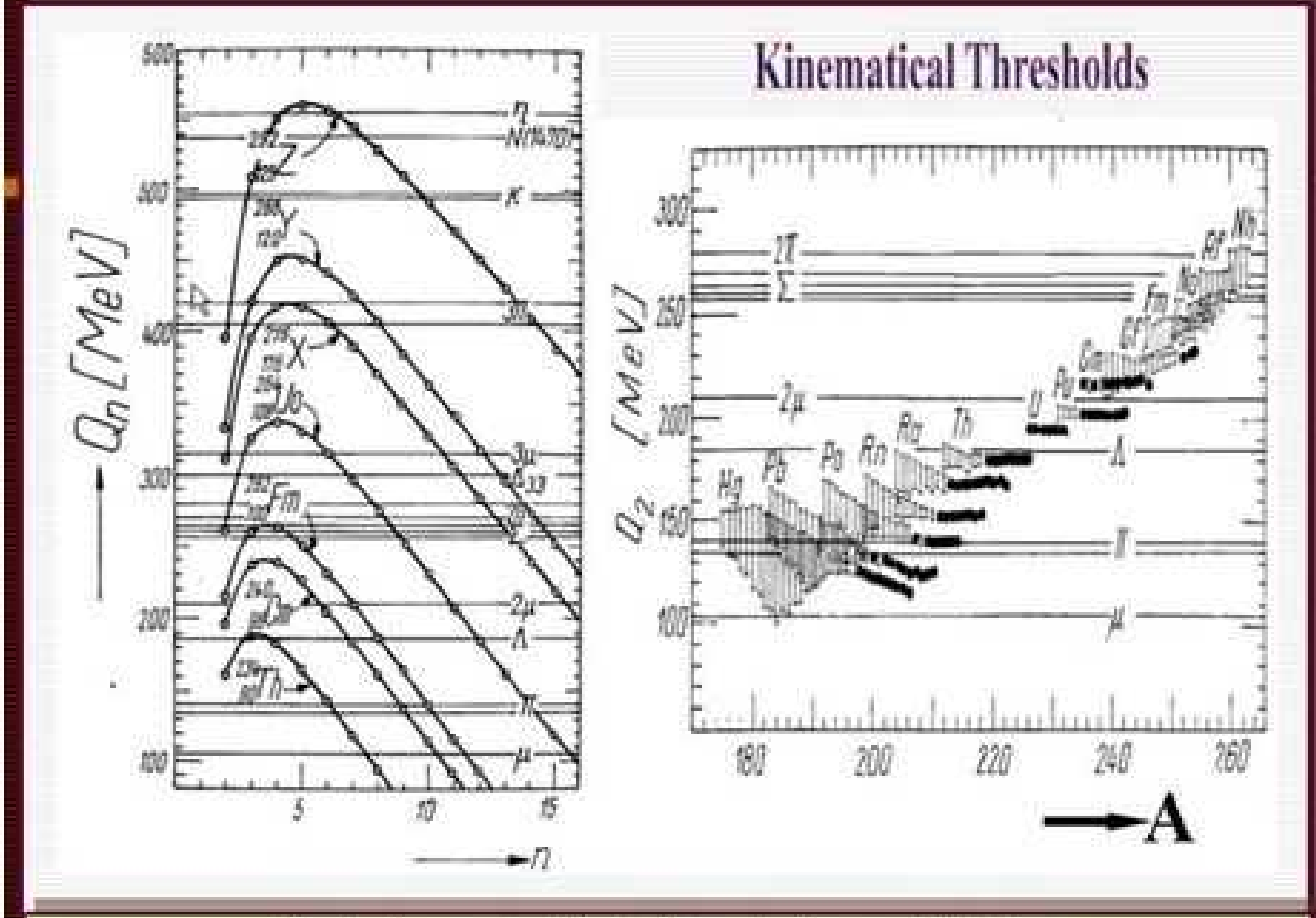

Fig.1a: Qn -energies liberated in a n-body symmetric fission

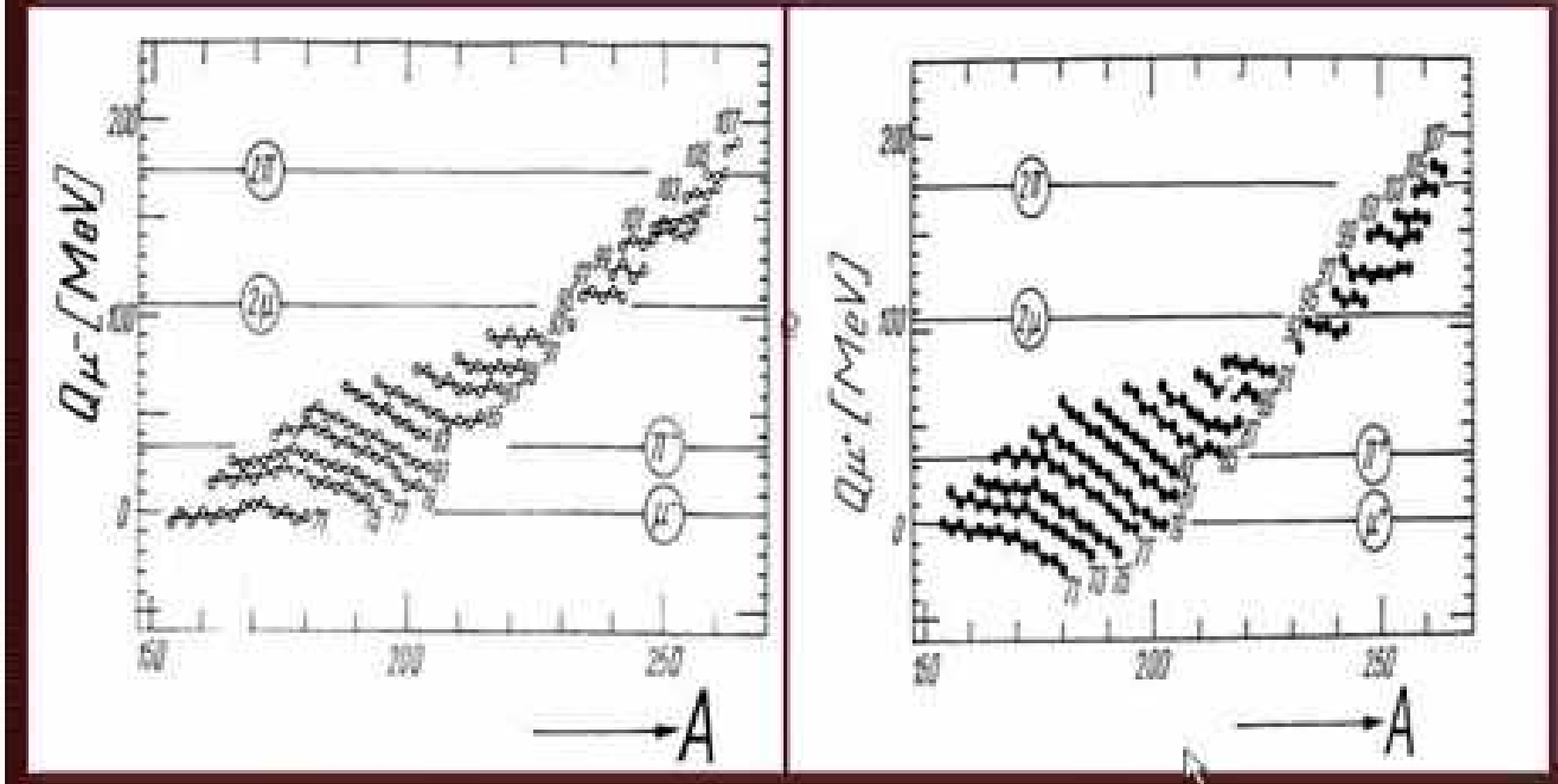

Fig.1b: Qmu-energies liberated in two-body symmetric muonic fission

In Fig.1a we presented the values of energies liberated in two-body symmetric fission as well as in n-body fission of heavy nuclei. The kinematical thresholds for different particle emission during fission are indicated by horizontal lines.

Therefore, from the results Fig.1a,b, we can see that different natural muonic radioactivities are energetically possible as follows:

$\mu^{\pm}-$natural radioactivities (promt muons) for parent nuclei with $Z>72$

$\pi^{\pm}-$natural radioactivities (delayed muons) for parent nuclei with $Z>76$

$2\mu^{\pm}-$natural radioactivities (promt muons) for parent nuclei with $Z>91$

$2\pi^{\pm}-$natural radioactivities (delayed muons) for parent nuclei with $Z>100$

## 2.2 Physical (A,Z)-regions where the parent nuclei are able to emit the $(\mu, \nu_{\mu})-$pair during the spontaneous fission.

These investigations was conducted in agreement with a special *fission-like model*, for muon emission during fission, similar with that presented in our paper [2] for the pion emission during fission. Hence, a fission-like model for the muonic radioactivity was also regarded as a first stage in the development of an approximate theory of this new phenomenon that takes into account the essential degree of freedom of the system: $\mu-$*fissility* , $\mu-$*fission barrier height*, etc.

Therefore, let us consider

$$E_C^{\mu F}(0)=E_C^0-\alpha(m_\mu+m_\nu) \tag{1}$$

$$E_S^{\mu F}(0)=E_S^0-(1-\alpha)(m_\mu+m_\nu) \tag{2}$$

where $E_C^0$ and $E_S^0$ are the usual Coulomb energy and surface energy of the parent nucleus given by

$$E_C^0=\gamma Z^2/A^{1/3} \text{ and } E_S^0=\beta A^{2/3} \tag{3}$$

with $\beta=17.80$ MeV and $\gamma=0.71$ . $\alpha$ is a parameter defined so that $\alpha m_\mu$ and $(1-\alpha)m_\mu$ are the Colombian and nuclear contributions to the muon mass (for $\alpha=1$ , the entire muon mass is obtained only from Coulomb energy of the parent nucleus). So, by analogy with binary fission, we introduced the muonic fissility parameters $X_{\mu F}^{\alpha}$ in Fig. 2 while the barrier height in terms of these parameters are given in Fig.3.

Therefore, in these $\mu$ -fission models we introduced the essential parameters of these new types of radioactivitis such as: $Q_{\mu F}$ -*values*, $\mu$ -*fissility parameters*, $\mu$ -*fission barrier* (or kinematical thresholds)(see Figs 1-3), nuclear configuration of the parent nucleus at μF-saddle points, etc.

Therefore, in Fig. 2 we presented the regions from the plane (A, Z) in which the parent nuclei are able to emit spontaneously muons during the nuclear spontaneous fission.

Therefore, according to Fig. 2, we have the following important regions

- SHE (super heavy elements)-region, indicated by white circles, where $X_{\mu F}>1$ and $Q_\mu>0$ all the nuclei are able to emit spontaneously the pair (muon, neutrino) during the SF since no muonic fission barrier exists.
- HE (heavy elements)-zone marked by signs plus (+++), corresponding to $X_{\mu F}<1$ and $Q_\mu>0$ , where all the nuclei can emit spontaneously the pair (muon, neutrino) only by quantum tunnelling of the muonic fission barrier.
- E-region, indicated by signs minus (----) where the spontaneous pair (muon, neutrino) emission is energetically interdicted since $Q_\mu<0$ .

# Physical Regions for Muonic Radioactivity

$$(A,Z) \rightarrow \mu^{+}(\mu^{-}) + \nu_{\mu}(\overline{\nu_{\mu}}) + (A_1,Z_1) + \ldots + (A_n,Z_n)$$

$(A,Z) \rightarrow \mu^{\pm} + anything$ **(muonic radioactivity (±))**

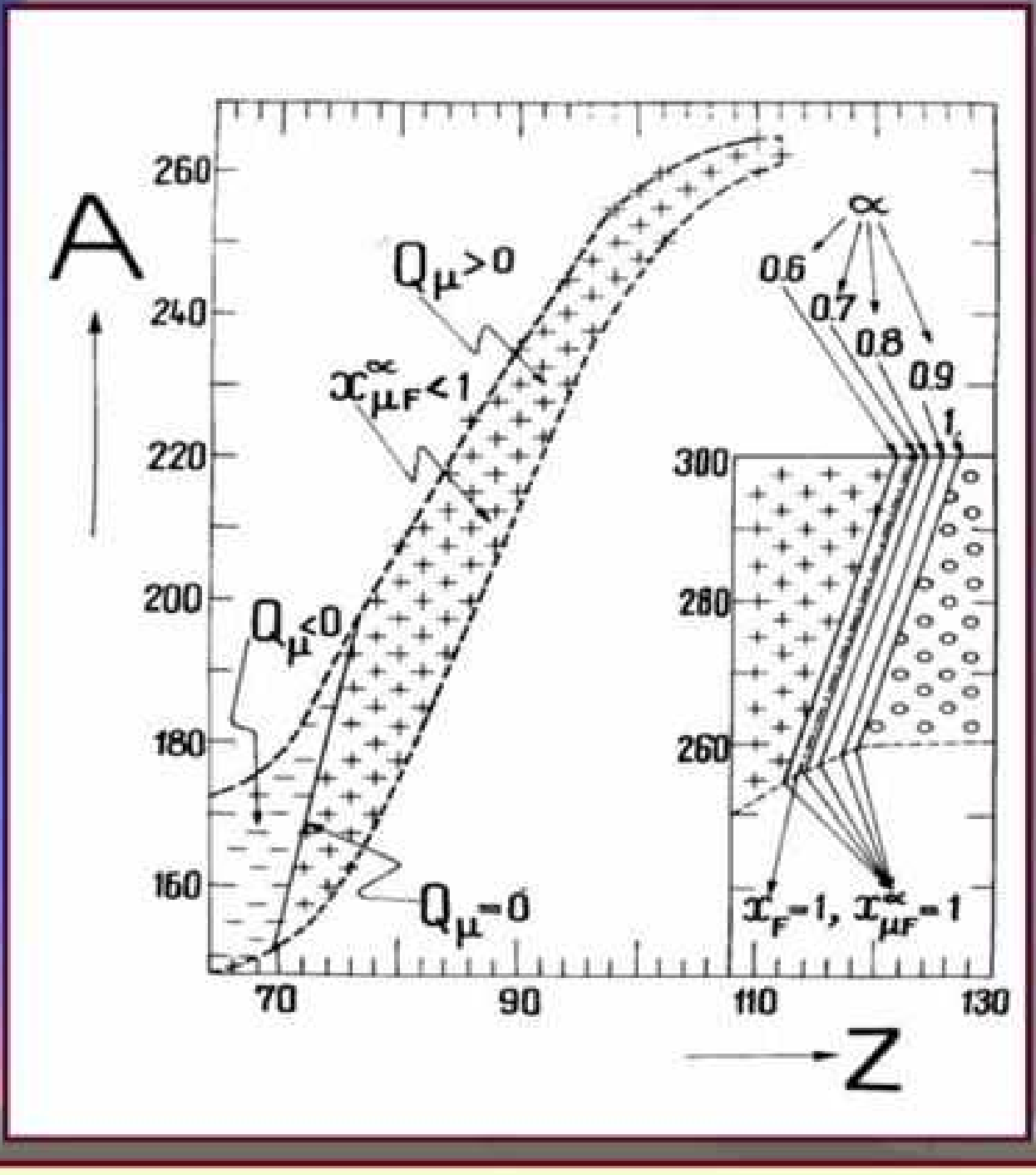

In a muonic fission-like model we defined

$$Q_{\mu n} = M_A - M_{A_1} - \ldots - M_{A_n} - m_\mu - m_\nu,$$

$\mu$-fissility parameter

$$X^\alpha_{\mu F} = \frac{E^\alpha_C(0)}{2E^\alpha_S(0)} = \frac{E^0_C - \alpha m_\mu}{2[E^0_S - (1-\alpha)m_\mu]},$$

$$\left(\frac{Z^2}{A}\right)_{\mu F} = \frac{Z^2}{A} - \frac{m_\mu}{\gamma A^{1/3}} \frac{\alpha - (1-\alpha)E^0_C/E^0_S}{1-(1-\alpha)m_\mu/E^0_S}$$

where:

$E^\alpha_C(0) = E^0_C - \alpha m_\mu$, $E^\alpha_S(0) = E^0_S - (1-\alpha)m_\mu$,

$E^0_C = \gamma Z^2/A^{1/3}$, $E^0_S = \beta A^{2/3}$,

$\gamma = 0.71\text{MeV}$, $\beta = 17.80\text{MeV}$, $0 \le \alpha \le 1$

The regions from plane (A,Z) where the parent nuclei are able to emit spontaneously muons by tuneling a mu-fission barrier (++) and without the mu-fission barrier (oo).

Fig. 2: Physical regions in the plane (A,Z) for muonic fission

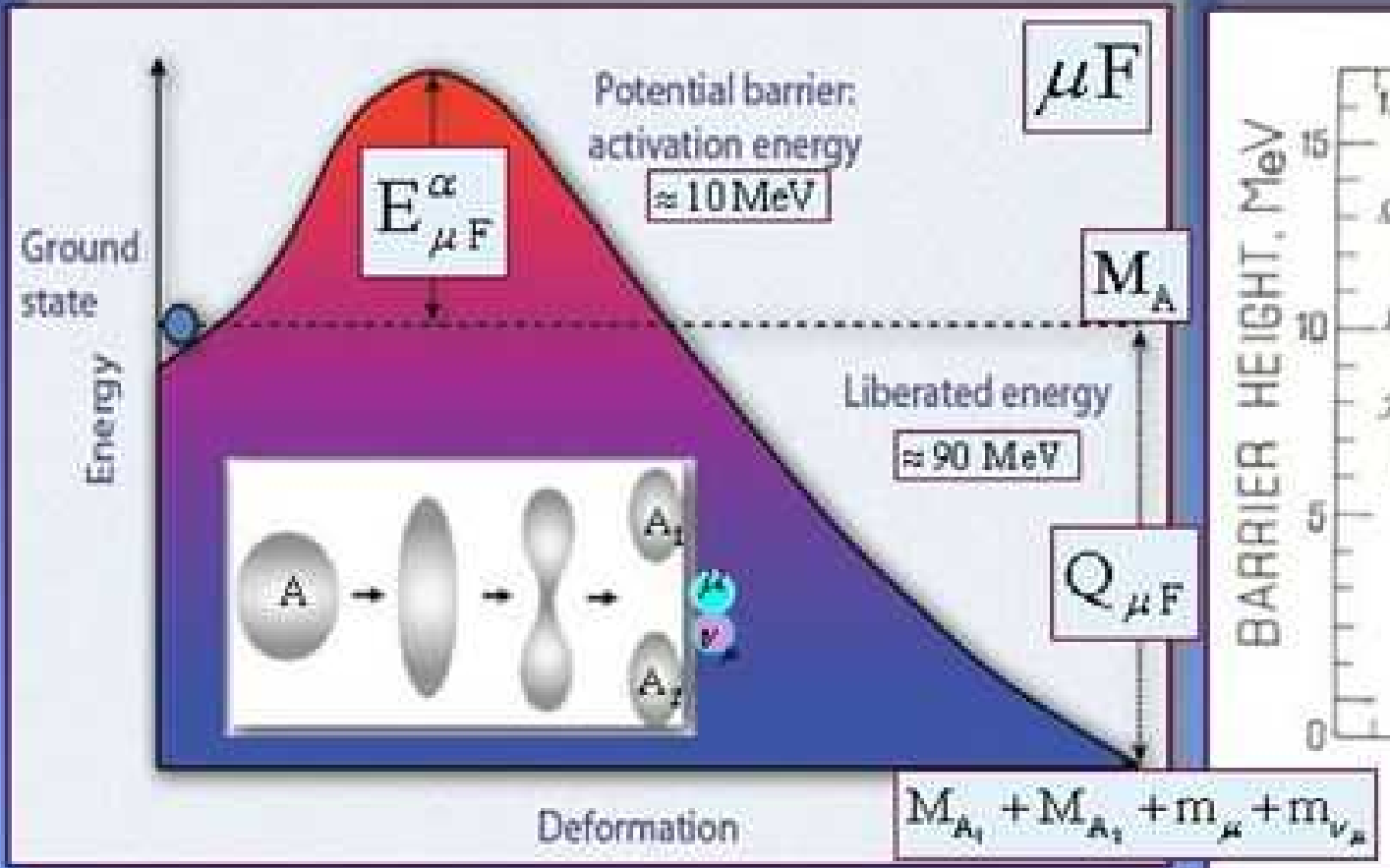

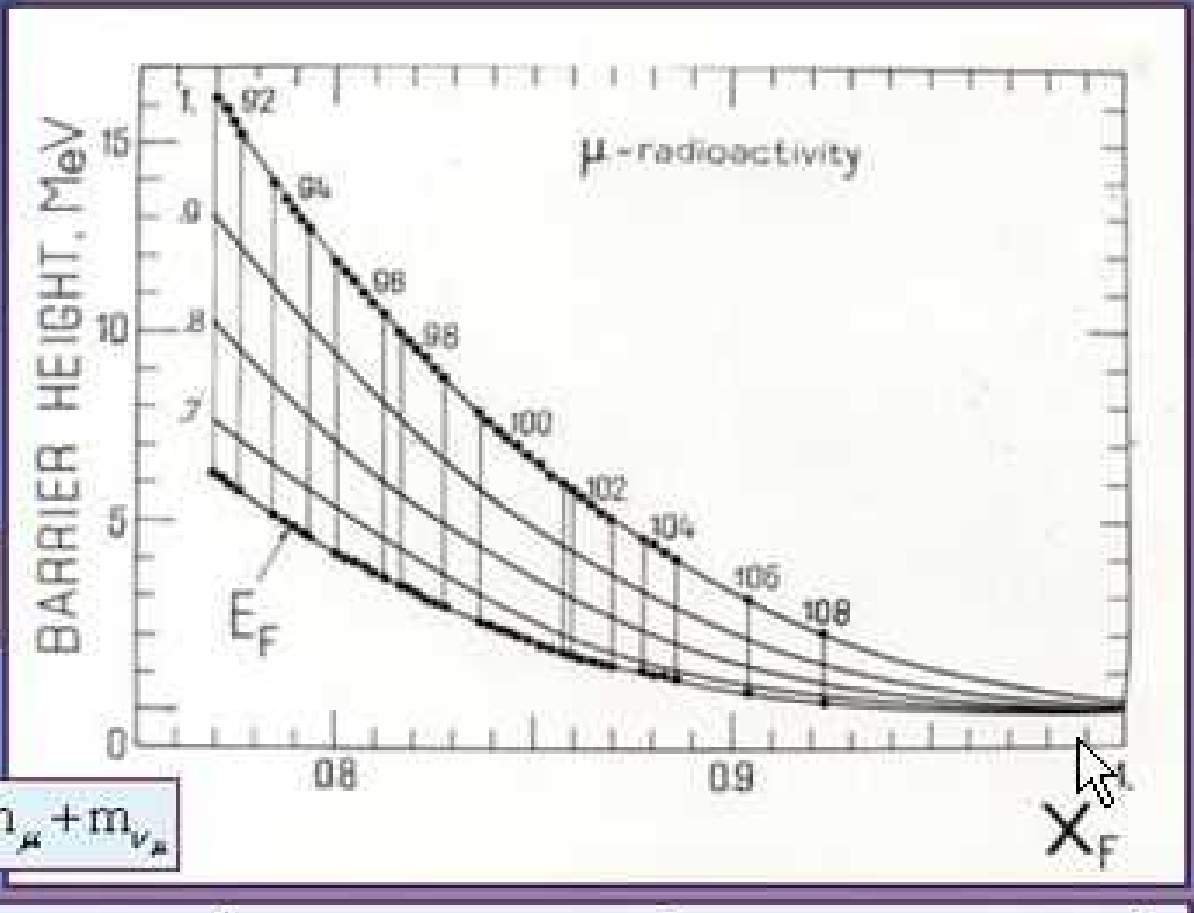

$$E^\alpha_{\mu F}(LD) = (E^\alpha_{\mu F})^0_S[\,0.73\,(1-X^\alpha_{\mu F})^3 - 0.33\,(1-X^\alpha_{\mu F})^4 + 1.92\,(1-X^\alpha_{\mu F})^5 - 0.21\,(1-X^\alpha_{\mu F})^6]$$

where :

$$(E^\alpha_{\mu F})^0_S = E^0_s - (1-\alpha)\,m_\mu\,,\ (E^\alpha_{\mu F})^0_C = E^0_C - \alpha\,m_\mu$$

and

$$X^\alpha_{\mu F} = [\,E^0_C - \alpha\,m_\mu]/2[\,E^0_S - (1-\alpha)\,m_\mu]$$

Fig. 3: Schematic description of muonic fission barrier (left) and liquid drop (LD)-model predictions for the barrier height of muon-neutrino emission during fission (right)

**2.3. Predictions for (muon-neutrino)-pair emission during the two-body spontaneous fission**

Next, the essential ingredients used for the estimation of the relative muonic yields $\Gamma_{\mu F}/\Gamma_{SF}$ are as follows:

(a) The variables $\theta_{SF}$ and $\theta_{\mu F}$ defined by

$$\theta_{SF}=\frac{Z^2}{A}-37.5 \tag{4}$$

$$\theta_{\mu F}^{\alpha}=\left(\frac{Z^2}{A}\right)_{\mu F}^{\alpha}-37.5=\theta-\frac{m_\mu}{\gamma A^{2/3}}\frac{\alpha-(1-\alpha)E_C^0/E_S^0}{1-(1-\alpha)m_\mu/E_S^0} \tag{5}$$

(b) The real scaling function $\tau(\theta)$, defined for spontaneous fission (SF) by Swiatecki [7] as follows

$$\tau(\theta_{SF})=\log \mathrm{T}_{\mathrm{xF}}(\theta_{SF})\ +(5-\theta_{SF})\delta M=a_0+a_1\theta_{SF}+\ldots, \tag{6}$$

where: $\delta M=[M_{\exp}-M(A,Z)]$, and $M(A,Z)$ is smooth reference mass of Green [8]

$$\tau(\theta)=\log_{10}T_{\mu F}+(5-\theta_{\mu F}^{\alpha})\delta M=a_0+a_1\theta_{\mu F}^{\alpha}+a_2\theta_{\mu F}^{\alpha\ 2}+\ldots \tag{7}$$

where $\delta M$ is just that defined in Eq.(6).

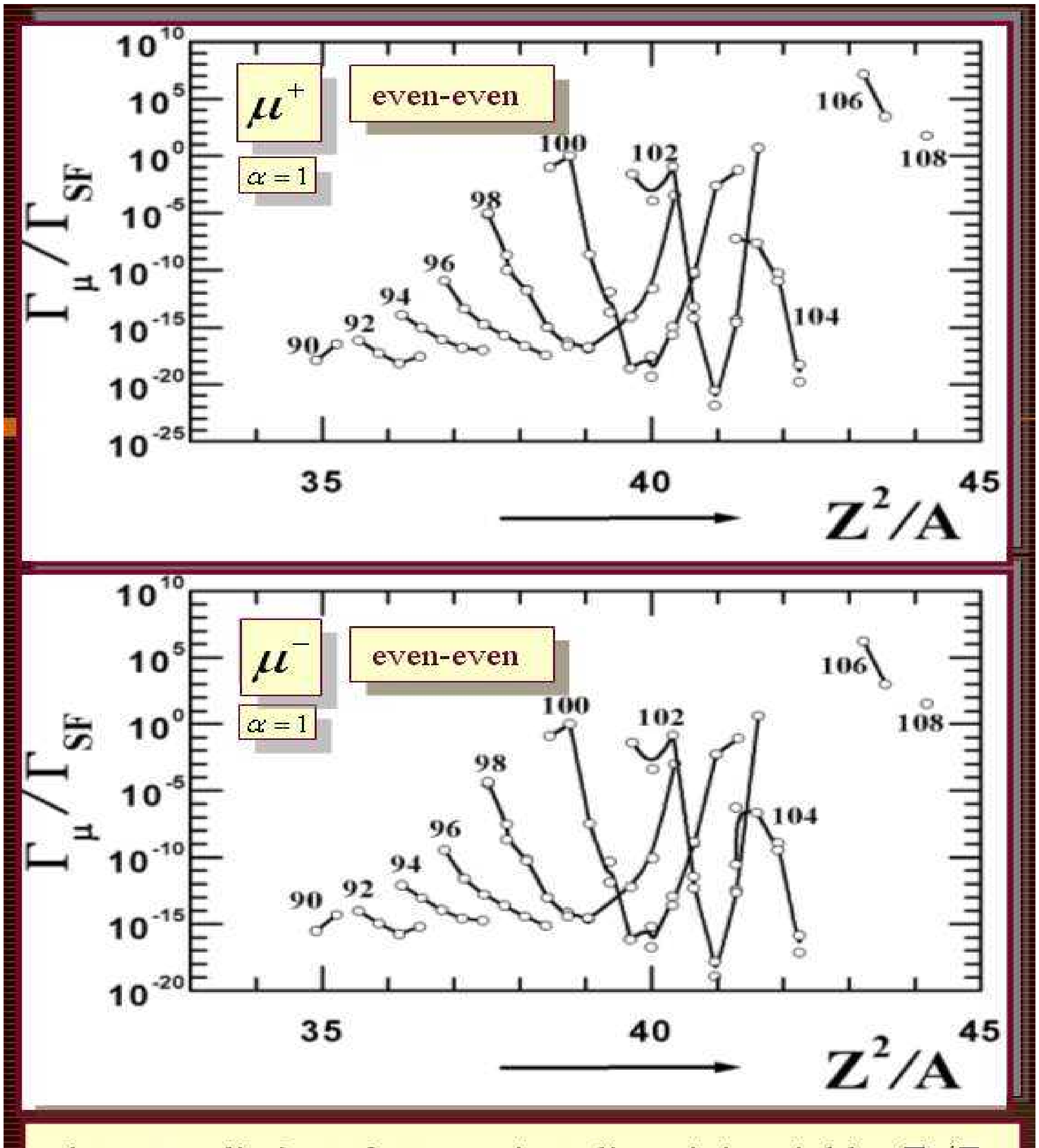

Fig.4: Predictions for muonic radioactivity yields: $\Gamma_\mu/\Gamma_{SF}$

Therefore, using Eqs. (6) and (7), with the constants $a_0$ and $a_1$ fitted with the spontaneous fission half lives data, we get $\tau(\theta)=19.70-6.13\theta$ for even-even parent nuclei as well $\tau(\theta)=28.21-7.32\theta$ for parent nuclei with A-odd . Then, we obtain the results presented in Fig. 4 for the muonic yields of even-even parent nuclei for $\alpha=1$ .

## 3. Experimental investigations of the spontaneous muon emission during fission.

**3.1. Results from LCP-identification emission spectrum [9]-[10]**.

The spontaneous muon emission as well as spontaneous pion emission during fission are processes which likened to the light charged particle emission (LCP) accompanying fission. Therefore, in order to obtain some useful informations on the existence of these new type of radioactivities we have reanalysed the available experimental results on the LCP emission from the spontaneous fission of $^{252}Cf$ and $^{259}Md$ and also on thermal neutron induced fission of $^{235}U$ (see our papers [9]-[10]). Here, in Figs. 5a-5b, we give only the results presented in Figs.2-3 of our paper [10]. Other results can be found in [10] (in Figs.4-5).

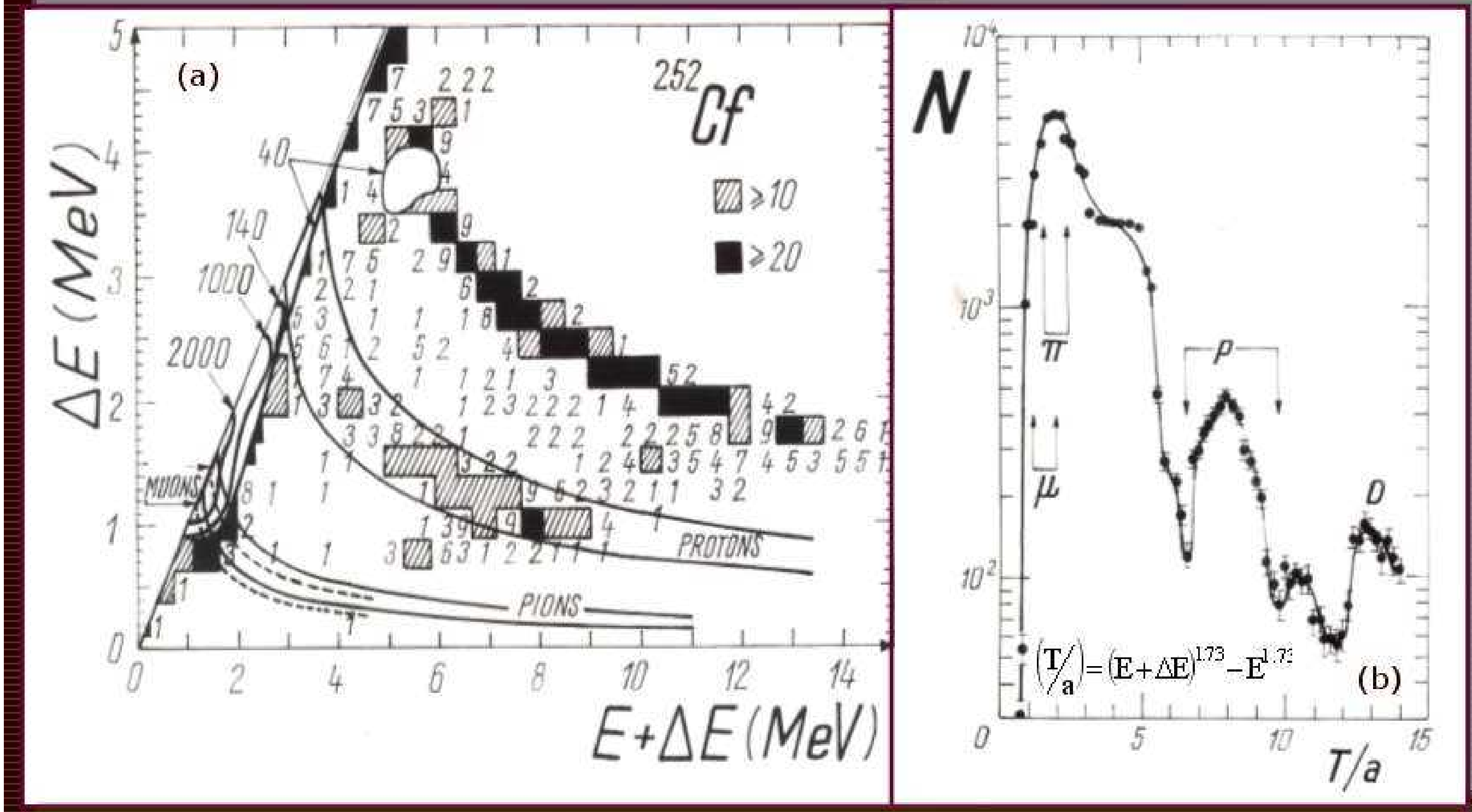

**Fig. 5a**: LCP-identification spectrum characteristics of light charged particles emissions during spontaneous fission of Cf-252.The experimental data are taken from S.L.Whetstone et al., Phys. Rev. 154, 1174 (1967) .

**Fig.5b**: The particle identification spectrum light charged particles emissions during spontaneous fission of Cf-252 calculated using the data from Fig.5a by summing all the events with the same particle identifiery output T/a .

Moreover, the unusual background, observed by J. F.Wild et al., [Phys. Rev. **C32,** 488 (1985)] in $(\Delta E, E+\Delta E)$ -energy region below that characteristic for long range alpha emission from $^{257}$Fm was interpreted by Ion, Bucurescu and Ion-Mihai [14] as being produced by negative pions emitted spontaneously by $^{257}$Fm. Then, they inferred value of the pionic plus muonic yield is: $\Gamma_{\pi+\mu}/\Gamma_{SF}<(1.2\pm0.2)\cdot10^{-3}$ $(\pi^-+\mu^-)$/fission. In a similar way, Janko and Povinec [16], obtained the yield: $\Gamma_{\pi+\mu}/\Gamma_{SF}<(7\pm6)\cdot10^{-5}$ $(\pi^++\mu^+)$/fission.

**3.2. Results using new spectrometer for investigation of ternary nuclear fission** [18].

Recently Khryachkov et al. [18] presented a new spectrometer for investigation of ternary nuclear fission. The measured characteristics of this spectrometer allow for its successful use in studies of ternary fission with the emission of $\alpha$-particles, tritons, and protons as well as in the search for exotic nuclear fission accompanied by the emission of charged mesons $(\pi^{\pm}, \mu^{\pm})$. This new spectrometer was tested with a reaction of spontaneous $^{252}$Cf ternary fission. This choice was determined by the fact that this reaction is well studied and the available data can be employed for the performance check of this facility. So, a $^{252}$Cf layer 5mm in diameter with an activity of 15 fissions/s was placed inside the spectrometer. The measurements were carried out continuously 1.5 months. A 2D spectrum of the scintillator signals obtained in coincidence with fragments is shown here in Fig.6 (see Fig. 9 in Ref. [18]).

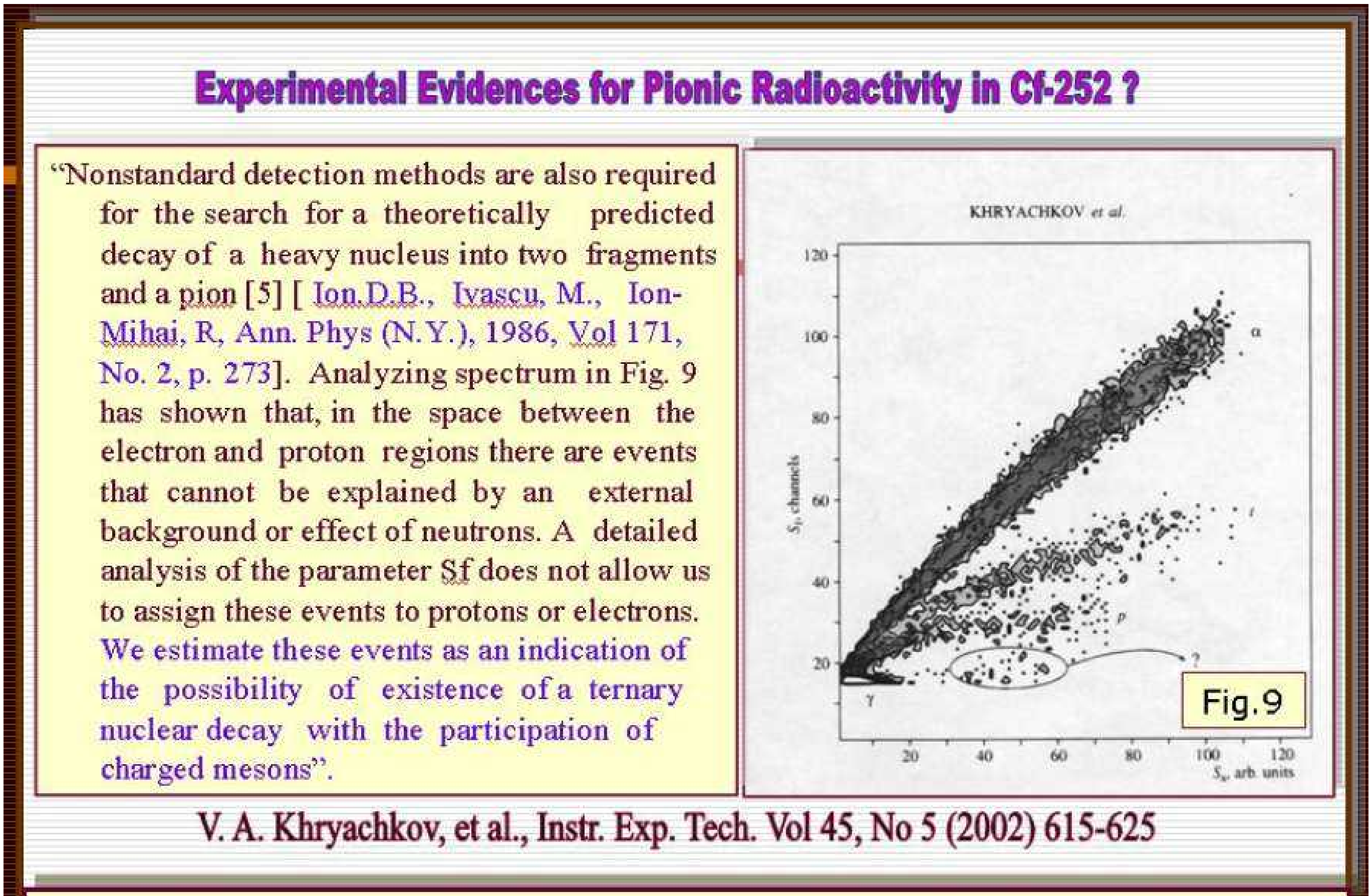

Fig.6: Experimental events (see above Fig.9) which are estimated to be indication for charged meson (pions and muons) emissions during spontaneous fission

**3.3. Supergiant Radioactive Halos as Integral Record of Muonic and Pionic Radioactivities [19].**

Pionic radiohalos (PIRH), as an integral record in time of the pionic nuclear radioactivity of the heavy nuclides with $Z > 80$ from the inclusions from ancient minerals, are introduced by us in Ref. [19] . It is then shown that the essential characteristic features of the supergiant halos (SGH) discovered by Grady and Walker [21] and Laemmlein [22] are reproduced with a surprisingly high accuracy by those of the pionic radiohalos. Indeed, the origin of radioactive halos was a mystery until the discovery of radioactivity and its power of coloration. The radioactive halos are of great interest for the nuclear physics because they are an integral record of radioactive decay in minerals. This integral record is detailed enough to allow estimation of the energy involved in the decay process and to identify the decaying nuclides through genetic connections. This latter possibility is exciting because there exist certain classes of halos, such as the *dwarf halos* [20], the X-*halos* [20], the *giant halos* [20] and the *supergiant halos* [21,22], which cannot be identified with the ring structure of the known alpha-emitters. Hence, barring the possibility of a non radioactive origin , these new variants of halos can be interpreted [20] as evidences for hitherto undiscovered alpha-radionuclide, as well as, signals for the existence of new types of radioactivities. In the spirit of these ideas, in our paper [19] we interpreted *supergiant radiohalos (SGH) as the integral record in time of the* $(\pi^{\pm}, \mu^{\pm})$ *-meson emission during fission of the heavy nuclides with Z>80 from the inclusions from ancient minerals* (see Ref.[19] for details).

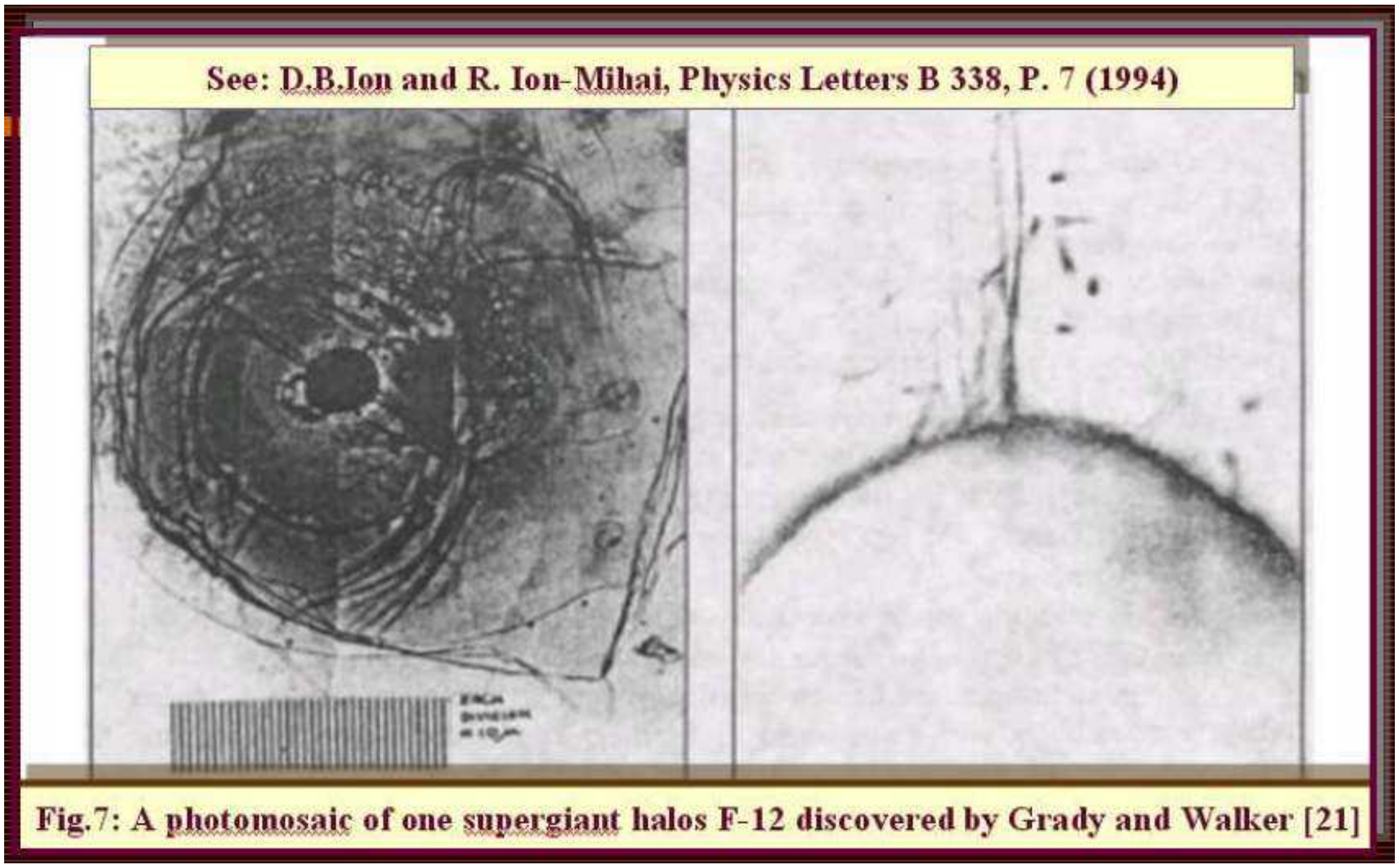

Fig.7: A photomosaic of one supergiant halos F-12 discovered by Grady and Walker [21]

### 3.4. The OPERA experiment for observation of spontaneous muon emission from Lead [23-24]

Natural lead is mainly composed by three nuclides: $^{206}\mathrm{Pb}(24.1\%)$ , $^{207}\mathrm{Pb}(22.1\%)$ and $^{208}\mathrm{Pb}(52.4\%)$ . They are stable nuclides, but the channels

(i) $(A,Z)\rightarrow\mu^{\pm}+\bar{\nu}_{\mu}(\nu_{\mu})+(A_1,Z_1)+...+(A_n,Z_n)$ (prompt muons)

(ii) $(A,Z)\rightarrow\pi^{\pm}(\pi^0)+(A_1,Z_1)+...+(A_n,Z_n)$ (delayed muons)

(i) and (ii) are energetically allowed (see Fig. 1). In the hypothesis of the decays (i) and (ii), the fission fragments would remain nearly at rest since the most of the available energy would be used to produce the μ (or $\pi$) and the kinetic energies of μ and $\nu_{\mu}$ (or $\pi$).

The spontaneous or neutron induced fission of Pb has never been observed. A search for a Pb muonic decay can be made as a byproduct of the OPERA experiment which aimed to confirm neutrino oscillations in the parameter region indicated by some atmospheric neutrino experiments. Therefore in ref. [23] it was proposed to perform an experimental search for muonic radioactivity from lead nuclei in the low background conditions offered by the Gran Sasso underground Laboratory (LNGS). The low cosmic muon flux and the low natural radioactivity of the rock in the experimental halls of the LNGS provide unique conditions, allowing a potential discovery, or, at least, to establish a good upper limit for this exotic decay process. A detailed description of the different background sources is given in ref. [23]. A test search for spontaneous emission of muons from Pb nuclei, using some OPERA lead/emulsion bricks, was described in [24]. They are in the process of making a complete simulation of the detector including its response and the track reconstruction efficiencies. They have shown that stringent limits for spontaneous muon radioactivity may be reached: $t_{1/2}=7\cdot 10^{23}$ years .

## 4. Summary and Conclusions

The experimental and theoretical results obtained on the muonic radioactivity in more than 25 years can be summarized as follows:

(i) We predicted that natural muonic radioactivities are energetically possible as follows (see Fig.1):

$$\mu^{\pm} - \text{natural radioactivities (promt muons) for parent nuclei with } Z > 72$$
$$\pi^{\pm} - \text{natural radioactivities (delayed muons) for parent nuclei with } Z > 76$$
$$2\mu^{\pm} - \text{natural radioactivities (promt muons) for parent nuclei with } Z > 91$$
$$2\pi^{\pm} - \text{natural radioactivities (delayed muons) for parent nuclei with } Z > 100$$

(ii) A fission-like model for the muonic radioactivity was regarded as a first stage in the development of an approximate theory of this new phenomenon that takes into account the essential degree of freedom of the system: $\mu - fissility$ , $\mu - fission$ *barrier height*, *nuclear configuration at* $\mu F -$ *saddle point,* etc. Detailed predictions for the (A,Z)-regions where parent nuclei are able to emit spontaneously prompt muons accompanied by two-body spontaneous are presented in Fig, 2.
(iii) The definitions of the $\mu - fission$ barrier height as well as their estimations via a liquid-drop model for some muonic-parent nuclei are displayed in Fig.3.
(iv) Numerical estimations on the $\Gamma_{\mu}/\Gamma_{SF}$ for the natural muonic radioactivities of the transuranium elements are presented in Fig.4. only for even-even parent nuclei.
(v) Experimental results from LCP-identification emission spectrum (see refs. [9]-[10]) are presented in Fig. 5. The unusual background, observed by Wild et al. in $(\Delta E, E+\Delta E)$-energy region below that characteristic for long range alpha emission from $^{257}$Fm was interpreted by us [14] as being produced by negative pions and negative muons emitted spontaneously in the spontaneous fission of $^{257}$Fm. Also, the results of Khryachkov et al [18], using new spectrometer for investigation of ternary nuclear fission, are included in Fig.6. They confirmed with accuracy our results presented in Fig.5.
(vi) The pionic $\pi^{\pm}-$ supergiant radiohalos introduced by us in the papers [19], can in fact, be considered as being produced by contribution of both kind of mesons (muons and pions). Then, the supergiant radiohalos (SGH), discovered by Grady and Walker [21] and Laemmlein [22] can be interpreted as being the $\pi^{-}+\mu^{-}-radiohalos$ and $\pi^{+}+\mu^{+}-radiohalos$ , respectively. Hence, these supergiant radiohalos (see Fig.7) can be considered as experimental evidences of the integral record in time of the natural $(\pi^{\pm}+\mu^{\pm})$ of the parent nuclei from radioactive inclusions present in ancient minerals.

Finally, we note that dedicated experiments, using parent nuclei with theoretically predicted high muonic and pionic yields (e.g. $^{258}$Fm, $^{259}$Fm, $^{258}$No, $^{260}$No, $^{254}$Rf, and $^{264}$Hs ), are desired since the discovery of the nuclear muonic radioactivity is not only of fundamental importance in nuclear science but also for the clarification of high instability of SHE- nuclei.
This research was supported by CNCSIS under contract ID-52-283/2007.